\documentclass[preprint*]{JHEP3} 

\usepackage{epsfig,multicol,bbm}

\newcommand\fverb{\setbox\pippobox=\hbox\bgroup\verb}
\newcommand\fverbdo{\egroup\medskip\noindent%
            \fbox{\unhbox\pippobox}\ }
\newcommand\fverbit{\egroup\item[\fbox{\unhbox\pippobox}]}
\newbox\pippobox

\newcommand{\be}{\begin{equation}}
\newcommand{\ee}{\end{equation}}
\newcommand{\bq}{\begin{eqnarray}}
\newcommand{\eq}{\end{eqnarray}}

\title{Statefinder diagnosis in a non-flat universe and the holographic model of dark energy}

\author{M. R. Setare\\
Department of Science, Payame Noor University, Bijar, Iran\\E-mail:
\email{rezakord@mail.ipm.ir}}
\author{Jingfei Zhang\\
    School of Physics and Optoelectronic
Technology, Dalian University of Technology, Dalian 116024, People's
Republic of China\\
    E-mail: \email{jfzhang@student.dlut.edu.cn}}
\author{Xin Zhang\\
    Institute of Theoretical Physics, Chinese
Academy of Sciences, P.O.Box 2735, Beijing 100080, People's Republic
of China\\
    Interdisciplinary Center of Theoretical Studies, Chinese
Academy of Sciences, P.O.Box 2735, Beijing 100080, People's Republic
of China\\
    E-mail: \email{zhangxin@itp.ac.cn}}

\abstract{In this paper, we study the holographic dark energy model
in non-flat universe from the statefinder viewpoint. We plot the
evolutionary trajectories of the holographic dark energy model for
different values of the parameter $c$ as well as for different
contributions of spatial curvature, in the statefinder
parameter-planes. The statefinder diagrams characterize the
properties of the holographic dark energy and show the
discrimination between this scenario and other dark energy models.
As we show, the contributions of the spatial curvature in the model
can be diagnosed out explicitly by the statefinder diagrams.
Furthermore, we also investigate the holographic dark energy model
in the $w-w'$ plane, which can provide us with a useful dynamical
diagnosis complement to the statefinder geometrical diagnosis.}

\begin{document}


Nowadays it is strongly believed that the universe is experiencing
an accelerated expansion. Recent observations from type Ia
supernovae (SNIa) \cite{SN} in associated with large scale structure
(LSS) \cite{LSS} and cosmic microwave background (CMB) anisotropies
\cite{CMB} have provided main evidence for this cosmic acceleration.
In order to explain why the cosmic acceleration happens, many
theories have been proposed. Although theories of trying to modify
Einstein equations constitute a big part of these attempts, the
mainstream explanation for this problem, however, is known as
theories of dark energy. It is the most accepted idea that a
mysterious dominant component, dark energy, with negative pressure,
leads to this cosmic acceleration, though its nature and
cosmological origin still remain enigmatic at present.

The combined analysis of cosmological observations suggests that the
universe consists of about $70\%$ dark energy, $30\%$ dust matter
(cold dark matter plus baryons), and negligible radiation. Although
the nature and origin of dark energy are unknown, we still can
propose some candidates to describe it. The most obvious theoretical
candidate of dark energy is the cosmological constant $\lambda$ (or
vacuum energy) \cite{Einstein:1917,cc} which has the equation of
state $w=-1$. However, as is well known, there are two difficulties
arise from the cosmological constant scenario, namely the two famous
cosmological constant problems --- the ``fine-tuning'' problem and
the ``cosmic coincidence'' problem \cite{coincidence}. The
fine-tuning problem asks why the vacuum energy density today is so
small compared to typical particle scales. The vacuum energy density
is of order $10^{-47} {\rm GeV}^4$, which appears to require the
introduction of a new mass scale 14 or so orders of magnitude
smaller than the electroweak scale. The second difficulty, the
cosmic coincidence problem, says: Since the energy densities of
vacuum energy and dark matter scale so differently during the
expansion history of the universe, why are they nearly equal today?
To get this coincidence, it appears that their ratio must be set to
a specific, infinitesimal value in the very early universe.

Theorists have made lots of efforts to try to resolve the
cosmological constant problem, but all these efforts were turned out
to be unsuccessful. However, there remain other candidates to
explaining dark energy. An alternative proposal for dark energy is
the dynamical dark energy scenario. The cosmological constant
puzzles may be better interpreted by assuming that the vacuum energy
is canceled to exactly zero by some unknown mechanism and
introducing a dark energy component with a dynamically variable
equation of state. The dynamical dark energy proposal is often
realized by some scalar field mechanism which suggests that the
energy form with negative pressure is provided by a scalar field
evolving down a proper potential. So far, a large class of
scalar-field dark energy models have been studied, including
quintessence \cite{quintessence}, $K$-essence \cite{kessence},
tachyon \cite{tachyon}, phantom \cite{phantom}, ghost condensate
\cite{ghost1,ghost2} and quintom \cite{quintom}, and so forth. But
we should note that the mainstream viewpoint regards the scalar
field dark energy models as an effective description of an
underlying theory of dark energy. In addition, other proposals on
dark energy include interacting dark energy models \cite{intde},
braneworld models \cite{brane}, and Chaplygin gas models \cite{cg},
etc.. One should realize, nevertheless, that almost these models are
settled at the phenomenological level, lacking theoretical root.

In recent years, many string theorists have devoted to understand
and shed light on the cosmological constant or dark energy within
the string framework. The famous Kachru-Kallosh-Linde-Trivedi (KKLT)
model \cite{kklt} is a typical example, which tries to construct
metastable de Sitter vacua in the light of type IIB string theory.
Furthermore, string landscape idea \cite{landscape} has been
proposed for shedding light on the cosmological constant problem
based upon the anthropic principle and multiverse speculation.
Another way of endeavoring to probe the nature of dark energy within
the fundamental theory framework originates from some considerations
of the features of the quantum gravity theory. It is generally
believed by theorists that we can not entirely understand the nature
of dark energy before a complete theory of quantum gravity is
established \cite{Witten:2000zk}. However, although we are lacking a
quantum gravity theory today, we still can make some attempts to
probe the nature of dark energy according to some principles of
quantum gravity. The holographic dark energy model is just an
appropriate example, which is constructed in the light of the
holographic principle of quantum gravity theory. That is to say, the
holographic dark energy model possesses some significant features of
an underlying theory of dark energy.

The distinctive feature of the cosmological constant or vacuum
energy is that its equation of state is always exactly equal to
$-1$. However, when considering the requirement of the holographic
principle originating from the quantum gravity speculation, the
vacuum energy will acquire dynamically property. As we speculate,
the dark energy problem may be in essence a problem belongs to
quantum gravity \cite{Witten:2000zk}. In the classical gravity
theory, one can always introduce a cosmological constant to make the
dark energy density be an arbitrary value. However, a complete
theory of quantum gravity should be capable of making the properties
of dark energy, such as the energy density and the equation of
state, be determined definitely and uniquely. Currently, an
interesting attempt for probing the nature of dark energy within the
framework of quantum gravity is the so-called ``holographic dark
energy'' proposal
\cite{Cohen:1998zx,Horava:2000tb,Hsu:2004ri,Li:2004rb}. It is well
known that the holographic principle is an important result of the
recent researches for exploring the quantum gravity (or string
theory) \cite{holoprin}. This principle is enlightened by
investigations of the quantum property of black holes. Roughly
speaking, in a quantum gravity system, the conventional local
quantum field theory will break down. The reason is rather simple:
For a quantum gravity system, the conventional local quantum field
theory contains too many degrees of freedom, and such many degrees
of freedom will lead to the formation of black hole so as to break
the effectiveness of the quantum field theory.

For an effective field theory in a box of size $L$, with UV cut-off
$\Lambda$ the entropy $S$ scales extensively, $S\sim L^3\Lambda^3$.
However, the peculiar thermodynamics of black hole \cite{bh} has led
Bekenstein to postulate that the maximum entropy in a box of volume
$L^3$ behaves nonextensively, growing only as the area of the box,
i.e. there is a so-called Bekenstein entropy bound, $S\leq
S_{BH}\equiv\pi M_P^2L^2$. This nonextensive scaling suggests that
quantum field theory breaks down in large volume. To reconcile this
breakdown with the success of local quantum field theory in
describing observed particle phenomenology, Cohen et al.
\cite{Cohen:1998zx} proposed a more restrictive bound -- the energy
bound. They pointed out that in quantum field theory a short
distance (UV) cut-off is related to a long distance (IR) cut-off due
to the limit set by forming a black hole. In other words, if the
quantum zero-point energy density $\rho_{\Lambda}$ is relevant to a
UV cut-off $\Lambda$, the total energy of the whole system with size
$L$ should not exceed the mass of a black hole of the same size,
thus we have $L^3\rho_{\Lambda}\leq LM_P^2$. This means that the
maximum entropy is in order of $S_{BH}^{3/4}$. When we take the
whole universe into account, the vacuum energy related to this
holographic principle \cite{holoprin} is viewed as dark energy,
usually dubbed holographic dark energy. The largest IR cut-off $L$
is chosen by saturating the inequality so that we get the
holographic dark energy density
\begin{equation}
\rho_{\Lambda}=3c^2M_P^2L^{-2}~,\label{de}
\end{equation} where $c$ is a numerical constant, and $M_P\equiv 1/\sqrt{8\pi
G}$ is the reduced Planck mass. Many authors have devoted to
developed the idea of the holographic dark energy. It has been
demonstrated that it seems most likely that the IR cutoff is
relevant to the future event horizon
\begin{equation}
R_{\rm h}(a)=a\int_t^\infty{dt'\over a(t')}=a\int_a^\infty{da'\over
Ha'^2}~.\label{eh}
\end{equation} Such a holographic dark energy looks reasonable, since
it may provide simultaneously natural solutions to both dark energy
problems as demonstrated in Ref.\cite{Li:2004rb}. The holographic
dark energy model has been tested and constrained by various
astronomical observations \cite{obs1,obs2,obs3}. Furthermore, the
holographic dark energy model has been extended to include the
spatial curvature contribution, i.e. the holographic dark energy
model in non-flat space \cite{nonflat}. We focus in this paper on
the holographic dark energy in a non-flat universe. For other
extensive studies, see e.g. \cite{holoext}.

On the other hand, since more and more dark energy models have been
constructed for interpreting or describing the cosmic acceleration,
the problem of discriminating between the various contenders is
becoming emergent. In order to be capable of differentiating between
those competing cosmological scenarios involving dark energy, a
sensitive and robust diagnosis for dark energy models is a must. In
addition, for some geometrical models arising from modifications to
the gravitational sector of the theory, the equation of state no
longer plays the role of a fundamental physical quantity, so it
would be very useful if we could supplement it with a diagnosis
which could unambiguously probe the properties of all classes of
dark energy models. For this purpose a diagnostic proposal that
makes use of parameter pair $\{r,s\}$, the so-called ``statefinder",
was introduced by Sahni et al. \cite{sahni}. The statefinder probes
the expansion dynamics of the universe through higher derivatives of
the expansion factor $\stackrel{...}{a}$ and is a natural companion
to the deceleration parameter $q$ which depends upon $\ddot a$. The
statefinder is a ``geometrical'' diagnosis in the sense that it
depends upon the expansion factor and hence upon the metric
describing space-time.

In this paper we apply the statefinder diagnosis to the holographic
dark energy model in the non-flat universe. The statefinder can also
be used to diagnose different cases of the model, including various
model parameters and different spatial curvature contributions.
Analysis of the observational data provides constraints on the
holographic dark energy model. In \cite{obs1,obs2}, it has been
shown that regarding the observational data including type Ia
supernovae, cosmic microwave background, baryon acoustic
oscillation, and the X-ray gas mass fraction of galaxy clusters, the
holographic dark energy in flat universe behaves like a quintom-type
dark energy. Moreover, in \cite{obsnonflat}, it has been shown that
when including the spatial curvature contribution, the closed
universe and quintom-type dark energy are marginally favored, in the
light of SNIa and CMB data. We use the statefinder to diagnose these
different cases, and the result shows that the information of the
spatial curvature can be precisely diagnosed in the statefinder
planes.


Consider now the homogenous and isotropic universe described by the
Friedmann-Robertson-Walker (FRW) metric \be\label{metr}
ds^{2}=-dt^{2}+a^{2}(t)(\frac{dr^2}{1-kr^2}+r^2d\Omega^{2}), \ee
where $k$ denotes the curvature of the space with $k=0$, 1 and $-1$
corresponding to flat, closed and open universes, respectively. The
IR cutoff of the universe in the holographic model $L$ is defined as
\be L=a(t)r, \ee where $r$ is relevant to the future event horizon
of the universe. Given the fact that
\begin{eqnarray}
\int_0^{r_1}{dr\over \sqrt{1-kr^2}}&=&\frac{1}{\sqrt{|k|}}{\rm
sinn}^{-1}(\sqrt{|k|}\,r_1)\nonumber\\
&=&\left\{\begin{array}{ll}
\sin^{-1}(\sqrt{|k|}\,r_1)/\sqrt{|k|},\ \ \ \ \ \ &k=1,\\
r_1,&k=0,\\
\sinh^{-1}(\sqrt{|k|}\,r_1)/\sqrt{|k|},&k=-1,
\end{array}\right.
\end{eqnarray}
one can easily derive \be \label{leh} L=\frac{a(t) {\rm
sinn}[\sqrt{|k|}\,R_{h}(t)/a(t)]}{\sqrt{|k|}},\ee where $R_{\rm h}$
is the future event horizon given by (\ref{eh}). With normal
pressureless matter and holographic dark energy as sources, the
Friedmann equations take the form
 \bq \label{friedmann}
H^2=\frac{8\pi G}{3}\rho -\frac{k}{a^2}, \\ \nonumber \frac{\ddot
a}{a}=-\frac{4\pi G}{3}(\rho+3P),
 \eq
where $\rho$ is the total energy density, $\rho=\rho_{\rm m} +
\rho_\Lambda$, and $P=w\rho_\Lambda$ is the pressure of the dark
energy component since we know that the normal dust matter is
pressureless. Define the density parameters as usual
\begin{equation} \label{2eq9} \Omega_{\rm
m}=\frac{\rho_{\rm m}}{\rho_{\rm cr}}=\frac{\Omega_{\rm
m0}H^2_0}{H^2a^3},\hspace{0.5cm}\Omega_{
\Lambda}=\frac{\rho_{\Lambda}}{\rho_{\rm cr}}=\frac{c^2}{L^2
H^2},\hspace{0.5cm}\Omega_{k}=\frac{k}{a^2H^2}=\frac{\Omega_{k0}H^2_0}{a^2H^2},
\end{equation}
then we can rewrite the first Friedmann equation as
\begin{equation} \label{2eq10} \Omega_{\rm m}+\Omega_{\rm
\Lambda}=1+\Omega_{k}.
\end{equation}
Since we have \be\frac{\Omega_k}{\Omega_{\rm
m}}=a\frac{\Omega_{k0}}{\Omega_{\rm m0}}=a\gamma,\ee where
$\gamma=\Omega_{k0}/\Omega_{\rm m0}$, we get $\Omega_k=\Omega_{\rm
m} a\gamma$ and \be \label{wmeq} \Omega_{\rm
m}=\frac{1-\Omega_\Lambda}{1-a\gamma}. \ee Hence, from the above
equation, we get \be \label{wmeq1}
\frac{1}{aH}=\frac{1}{H_0}\sqrt{\frac{a(1-\Omega_\Lambda)}{\Omega_{\rm
m0}(1-a\gamma)}}. \ee Combining Eqs. (\ref{leh}) and (\ref{wmeq1}),
and using the definition of $\Omega_\Lambda$, we obtain
\begin{eqnarray} \label{wleq} \sqrt{|k|}\frac{R_{h}}{a}&=&{\rm
sinn}^{-1}\left[c\sqrt{|\gamma|}\sqrt{\frac{a(1-\Omega_\Lambda)}{\Omega_\Lambda(1-a\gamma)}}\,\right]\nonumber\\
&=&{\rm sinn}^{-1}(c\sqrt{|\Omega_k|/\Omega_\Lambda}).
\end{eqnarray}
From the above relationships, the differential equation describing
the evolutionary behavior of the holographic dark energy can be
derived \cite{nonflat,obsnonflat}
\begin{equation}
(1+z){d\Omega_\Lambda\over dz}=-{2\over
c}\Omega_\Lambda^{3/2}(1-\Omega_\Lambda)\sqrt{1-{c^2\gamma(1-\Omega_\Lambda)\over
\Omega_\Lambda(1+z-\gamma)}}-(1+z){\Omega_\Lambda(1-\Omega_\Lambda)\over
1+z-\gamma},\label{deq}
\end{equation}
where $z=(1/a)-1$ is the red-shift parameter of the universe. This
equation completely describes the dynamical evolution of the
holographic dark energy, and it can be solved numerically. From the
energy conservation equation of the dark energy, the equation of
state of dark energy can be given \cite{obsnonflat}
\begin{eqnarray} \label{ehol}
w
&=&-1-\frac{1}{3}\frac{d \ln \rho_\Lambda}{d \ln a}\nonumber\\
&=&-\frac{1}{3}\left[1+\frac{2}{c}\sqrt{\Omega_\Lambda}{\rm
cosn}(\sqrt{|k|}\,R_{h}/a)\right]\nonumber\\
&=&-\frac{1}{3}\left[1+\frac{2}{c}\sqrt{\Omega_\Lambda-c^2\Omega_k}\right],
\end{eqnarray}
where
\begin{equation}
\frac{1}{\sqrt{|k|}}{\rm cosn}(\sqrt{|k|}x)
=\left\{\begin{array}{ll}
\cos(x),\ \ \ \ \ \ &k=1,\\
1,&k=0,\\
\cosh(x),&k=-1.
\end{array}\right.
\end{equation}

The holographic dark energy model has been tested and constrained by
various astronomical observations, in both flat and non-flat cases.
These observational data include type Ia supernovae, cosmic
microwave background, baryon acoustic oscillation, and the X-ray gas
mass fraction of galaxy clusters. According to the analysis of the
observational data for the holographic dark energy model, we find
that generally $c<1$, and the holographic dark energy thus behaves
like a quintom-type dark energy. When including the spatial
curvature contribution, the fitting result shows that the closed
universe is marginally favored. Here we summarized the main
constraint results as follows:

\begin{enumerate}

\item For flat universe, using only the SNIa data to constrain the holographic
dark energy model, we get the fit results: $c=0.21^{+0.41}_{-0.12}$,
$\Omega_{\rm m0}=0.47^{+0.06}_{-0.15}$, with the minimal chi-square
corresponding to the best fit $\chi^2_{\rm min}=173.44$ \cite{obs1}.
In this fitting, The SNIa data used are 157 ``gold'' data listed in
Riess et al. \cite{Riess:2004nr} including 14 high redshift data
from the {\it Hubble Space Telescope} (HST)/Great Observatories
Origins Deep Survey (GOODS) program and previous data. Furthermore,
when combining the information from SNIa \cite{Riess:2004nr}, CMB
\cite{CMB} and LSS \cite{BAO}, the fitting for the holographic dark
energy model gives the parameter constraints in 1 $\sigma$:
$c=0.81^{+0.23}_{-0.16}$, $\Omega_{\rm m0}=0.28\pm 0.03$, with
$\chi_{\rm min}^2=176.67$ \cite{obs1}. In this joint analysis, the
SNIa data are still the 157 ``gold'' data \cite{Riess:2004nr}, the
CMB information comes from the measured value of the CMB shift
parameter $R$ given by \cite{CMB} $R\equiv \Omega_{\rm
m0}^{1/2}\int_0^{z_{\rm CMB}}dz'/E(z')=1.716\pm 0.062$, where
$z_{\rm CMB}=1089$ is the redshift of recombination and $E(z)\equiv
H(z)/H_0$, and the LSS information is provided by the baryon
acoustic oscillation (BAO) measurement \cite{BAO} $A\equiv
\Omega_{\rm m0}^{1/2} E(z_{\rm BAO})^{-1/3}[(1/z_{\rm
BAO})\int_0^{z_{\rm BAO}}dz'/E(z')]^{2/3}=0.469\pm 0.017$, where
$z_{\rm BAO}=0.35$.

\item Also for the flat case, the X-ray gas mass
fraction of rich clusters, as a function of redshift, has also been
used to constrain the holographic dark energy model \cite{obs2}. The
$f_{\rm gas}$ values are provided by {\it Chandra} observational
data, the X-ray gas mass fraction of 26 rich clusters, released by
Allen et al. \cite{Allen:2004cd}. The main results, i.e. the 1
$\sigma$ fit values for $c$ and $\Omega_{\rm m0}$ are:
$c=0.61^{+0.45}_{-0.21}$ and $\Omega_{\rm m0}=0.24^{+0.06}_{-0.05}$,
with the best-fit chi-square $\chi_{\rm min}^2=25.00$ \cite{obs2}.

\item For the non-flat universe, the authors of \cite{obsnonflat} used the data
coming from the SNIa and CMB to constrain the holographic dark
energy model, and got the 1 $\sigma$ fit results:
$c=0.84^{+0.16}_{-0.03}$, $\Omega_{\rm m0}=0.29^{+0.06}_{-0.08}$,
and $\Omega_{k0}=0.02\pm 0.10$, with the best-fit chi-square
$\chi_{\rm min}^2=176.12$. Also, in this analysis, the SNIa data
come from the 157 ``gold'' data in Riess et al. \cite{Riess:2004nr},
and the CMB information still comes from the measured value of the
CMB shift parameter $R$ \cite{CMB}.

\end{enumerate}


Now we switch to discussing the statefinder diagnosis of the
holographic dark energy model. For characterizing the expansion
history of the universe, one defines the geometric parameters
$H=\dot{a}/a$ and $q=-\ddot{a}/aH^2$, namely the Hubble parameter
and the deceleration parameter. It is clear that $\dot{a}>0$ means
the universe is undergoing an expansion and $\ddot{a}>0$ means the
universe is experiencing an accelerated expansion. From the cosmic
acceleration, $q<0$, one infers that there may exist dark energy
with negative equation of state, $w<-1/3$ and likely $w\sim -1$, but
it is hard to deduce the information of the dynamical property of
$w$ from the value of $q$. In order to extract the information of
the dynamical evolution of $w$, it seems that we need the higher
time derivative of the scale factor, ${\stackrel{...}a}$. Another
motivation for proposing the statefinder parameters comes from the
merit that they can provide with a diagnosis which could
unambiguously probe the properties of all classes of dark energy
models including the cosmological models without dark energy
describing the cosmic acceleration. Though at present we can not
extract sufficiently accurate information of $\ddot{a}$ and
${\stackrel{...}a}$ from the observational data, we can expect,
however, the high-precision observations of next decade may be
capable of doing this. The statefinder parameters can be used to
diagnose the evolutionary behaviors of various cosmological models,
and discriminate them from each other. In what follows, we shall
exam the holographic dark energy model in non-flat universe using
the statefinder diagnosis.

Generically, the statefinder pair $\{r,s\}$ is defined as follows
\cite{Evans:2004iq}
 \bq \label{rs1}
 r\equiv\frac{\stackrel{...}a}{aH^3},~~~~~
 s\equiv\frac{r-\Omega_{\rm tot}}{3(q-\Omega_{\rm tot}/2)}.
 \eq
Here $\Omega_{\rm tot}$ is the total energy density, $\Omega_{\rm
tot}=\Omega_{\rm m}+\Omega_\Lambda=1+\Omega_k$. Note that the
parameter $r$ is also called cosmic jerk. Thus the set of quantities
describing the geometry is extended to include $\{H, q, r, s\}$.
Trajectories in the $s-r$ plane corresponding to different
cosmological models exhibit qualitatively different behaviors. The
spatially flat $\Lambda$CDM (cosmological constant $\lambda$ with
cold dark matter) scenario corresponds to a fixed point in the
diagram
\begin{equation}
\{s,r\}\bigg\vert_{\rm flat-\Lambda CDM} = \{ 0,1\} ~.\label{lcdm}
\end{equation}
Departure of a given dark energy model from this fixed point
provides a good way of establishing the ``distance'' of this model
from spatially flat $\Lambda$CDM \cite{sahni}. As demonstrated in
Refs.
\cite{Evans:2004iq,Alam:2003sc,Gorini:2002kf,Zimdahl:2003wg,Sahni:2004ai,
Zhang:2004gc,Zhang:2005xk,Zhang:2005rj,Zhang:2005yz,Dabrowski:2005fg,
Wu:2005ap,Hu:2005fu,Sahni:2006pa} the statefinder can successfully
differentiate between a wide variety of dark energy models including
the cosmological constant, quintessence, phantom, quintom, the
Chaplygin gas, braneworld models and interacting dark energy models.
We can clearly identify the ``distance'' from a given dark energy
model to the flat-$\Lambda$CDM scenatio by using the $r(s)$
evolution diagram. The current location of the parameters $s$ and
$r$ in these diagrams can be calculated in models. The current
values of $s$ and $r$ are evidently valuable since we expect that
they can be extracted from data coming from SNAP (SuperNovae
Acceleration Probe) type experiments. Therefore, the statefinder
diagnosis combined with future SNAP observations may possibly be
used to discriminate between different dark energy models. It is
notable that in the non-flat universe the $\Lambda$CDM model no
longer corresponds a fixed point in the statefinder plane, it
exhibits an evolutionary trajectory
\begin{equation}
\{s,r\}\bigg\vert_{\rm nonflat-\Lambda CDM} = \{ 0,\Omega_{\rm
tot}\} ~.\label{lcdm}
\end{equation}

\begin{figure}[htbp]
\centering
\begin{center}
\includegraphics[scale=0.8]{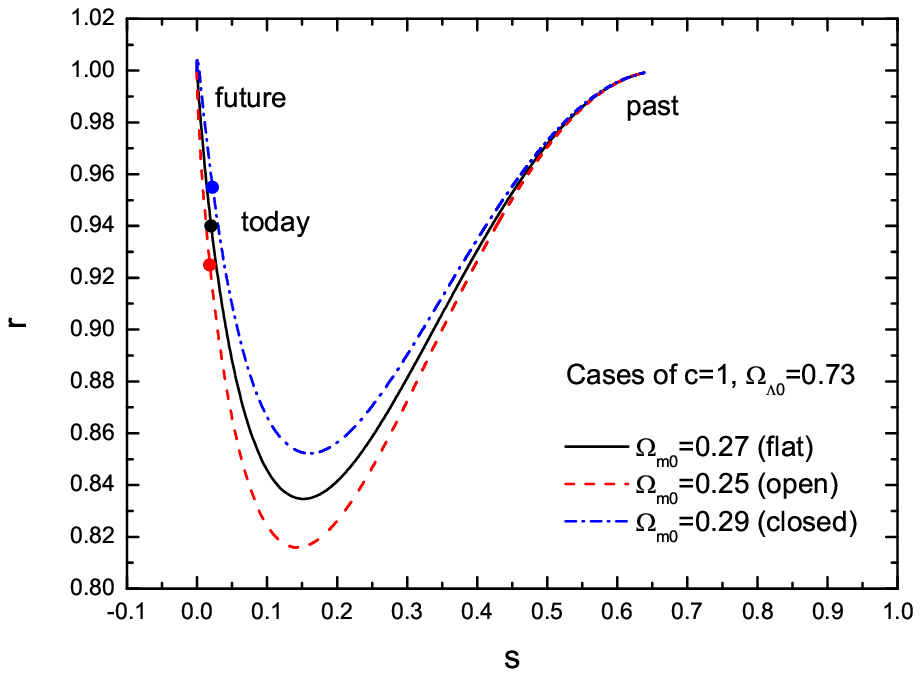}
$\begin{array}{c@{\hspace{0.2in}}c} \multicolumn{1}{l}{\mbox{}} &
\multicolumn{1}{l}{\mbox{}} \\
\includegraphics[scale=0.8]{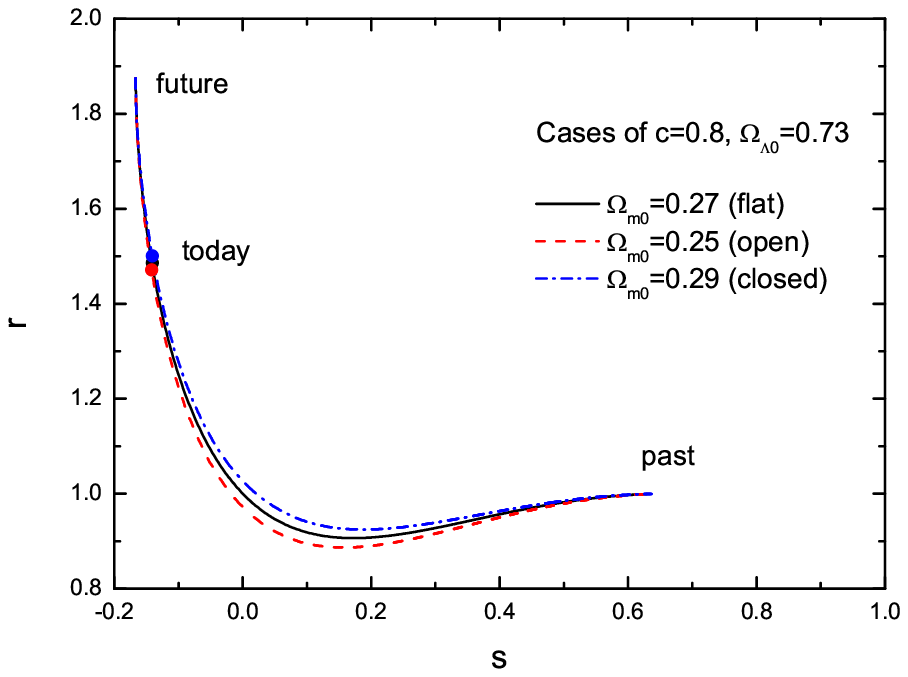} &\includegraphics[scale=0.8]{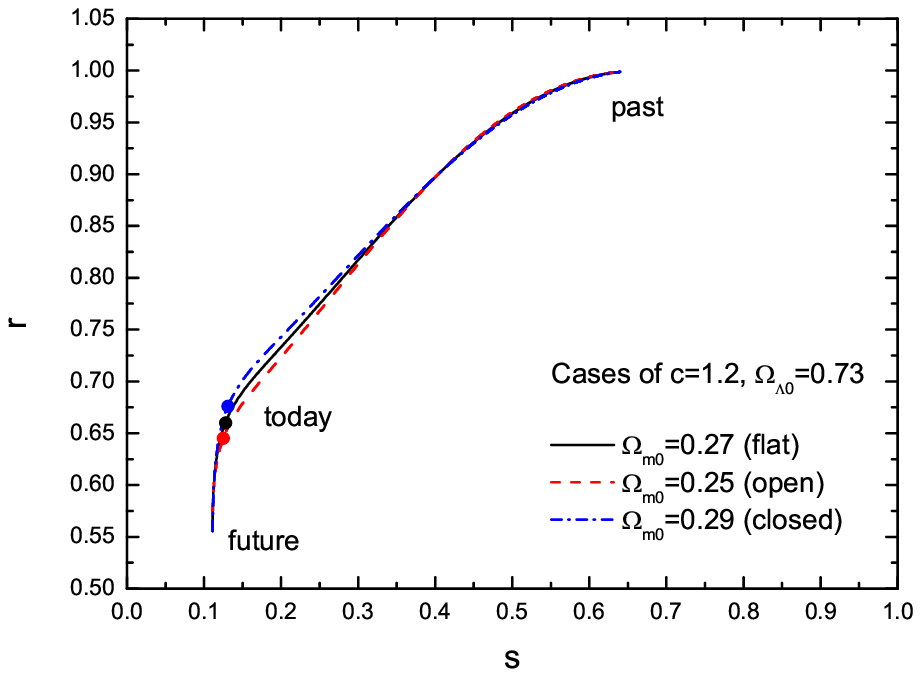} \\
\end{array}$
\end{center}
\caption[]{\small An illustrative example for the statefinder
diagnosis of the holographic dark energy model. We plot the
statefinder diagrams in the $s-r$ plane for the cases $c=1$, $c=0.8$
and $c=1.2$, respectively. Selected curves of $r(s)$ are plotted by
fixing $\Omega_{\Lambda 0}=0.73$ and varying $\Omega_{\rm m0}$ as
0.27, 0.25 and 0.29 corresponding to the flat, open and closed
universes, respectively. Dots locate the today's values of the
statefinder parameters $(s_0, r_0)$.} \label{fig:rseg}
\end{figure}

The statefinder parameters $r$ and $s$ can also be expressed as
\begin{equation}
r=\Omega_{\rm tot}+{9\over 2}w(1+w)\Omega_\Lambda-{3\over
2}w'\Omega_\Lambda~, \label{rw}
\end{equation}
\begin{equation}
s=1+w-{1\over 3}{w'\over w}~,  \label{sw}
\end{equation}
where the prime represents the derivative with respect to the
logarithm of the scale factor, $\ln a$. We also give the expression
of the deceleration parameter $q$,
\begin{equation}
q={1\over 2}\Omega_{\rm tot}+{3\over 2} w\Omega_\Lambda.
\end{equation}
It should be mentioned that the statefinder diagnosis for
holographic dark energy model in flat universe has been investigated
in detail in Ref.\cite{Zhang:2005yz}, where the focus is put on the
diagnosis of the different values of parameter $c$. In
Ref.\cite{Zhang:2005yz}, it has been demonstrated that from the
statefinder viewpoint $c$ plays a significant role in this model and
it leads to the values of $\{r, s\}$ in today and future
tremendously different. If the accurate information of $\{r_0,
s_0\}$ can be extracted from the future high-precision observational
data in a model-independent manner, these different features in this
model can be discriminated explicitly by experiments, one thus can
use this method to test the holographic dark energy model as well as
other dark energy models. Here we want to focus on the statefinder
diagnosis of the spatial curvature contribution in the holographic
dark energy model. The whole information of dynamics of this model
can be acquired from solving the differential equation (\ref{deq}).
Making the redshift $z$ vary in an enough large range involving far
future and far past, e.g. from $-1$ to order of several hundreds,
one can solve the differential equation (\ref{deq}) numerically and
then gets the evolution trajectories in the statefinder $s-r$ and
$q-r$ planes for this model. As an illustrative example, we plot the
statefinder diagrams in the $s-r$ plane for the cases $c=1$, $c=0.8$
and $c=1.2$, respectively, in figure \ref{fig:rseg}. Selected curves
of $r(s)$ are plotted by fixing $\Omega_{\Lambda 0}=0.73$ and
varying $\Omega_{\rm m0}$ as 0.27, 0.25 and 0.29 corresponding to
the flat, open and closed universes, respectively. Dots locate the
today's values of the statefinder parameters $(s_0, r_0)$. These
diagrams show that the evolution trajectories with different $c$
exhibit significantly different features in the statefinder plane,
which has been discussed in detail in Ref.\cite{Zhang:2005yz}. Now
we are interested in the diagnosis to the spatial curvature
contributions in the dark energy models using the statefinder
parameters as a probe. It is clearly shown in figure \ref{fig:rseg}
that the statefinder diagnosis has the power of testing the
contributions of spatial curvature.

\begin{figure}[htbp]
\centering $\begin{array}{c@{\hspace{0.2in}}c}
\multicolumn{1}{l}{\mbox{}} &
\multicolumn{1}{l}{\mbox{}} \\
\includegraphics[scale=0.8]{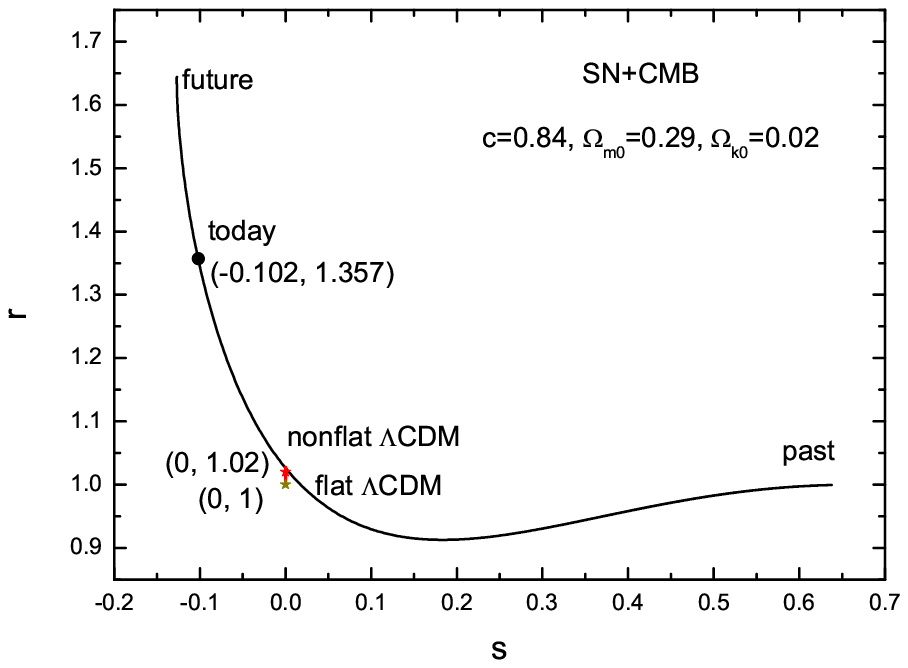} &\includegraphics[scale=0.8]{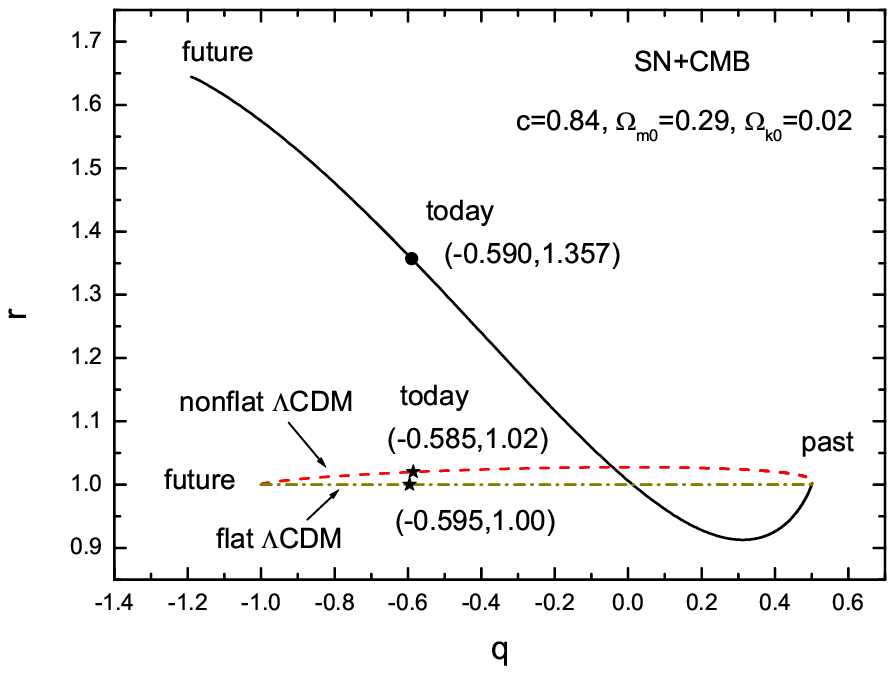} \\
\end{array}$
\caption[]{\small The statefinder diagrams $r(s)$ and $r(q)$ for the
spatially non-flat holographic dark energy model. The diagrams are
plotted in the light of the best-fit result of the SN+CMB data
analysis, $c=0.84$, $\Omega_{\rm m0}=0.29$, and $\Omega_{k0}=0.02$.
The coordinates of today locate at: $(s_0, r_0)=(-0.102, 1.357)$ and
$(q_0, r_0)=(-0.590, 1.357)$. For comparison, the statefinder
diagrams for $\Lambda$CDM model (in both spatially flat and non-flat
cases) are also plotted in the $s-r$ and $q-r$ planes. The
parameters for non-flat $\Lambda$CDM are also $\Omega_{\rm m0}=0.29$
and $\Omega_{k0}=0.02$, and the parameter for flat $\Lambda$CDM is
$\Omega_{\rm m0}=0.27$.} \label{fig:obs}
\end{figure}

On the other hand, associated with the current constraints for the
model from the observational data one can predict the statefinder
evolution trajectories $r(s)$ and $r(q)$ for the holographic dark
energy model. Due to the fact that the cases of flat universe have
been analyzed in detail in Ref.\cite{Zhang:2005yz}, we shall discuss
the case of non-flat universe here. As has been mentioned above, the
analysis of the current observational data involving the information
of SNIa and CMB gives the 1-$\sigma$ fit values of the parameters:
$c=0.84^{+0.16}_{-0.03}$, $\Omega_{\rm m0}=0.29^{+0.06}_{-0.08}$,
and $\Omega_{k0}=0.02\pm 0.10$. We should have analyzed the
cosmological evolution of the model with errors of confidence level
by means of the statefinder parameters, however, it can be seen
clearly that the errors for $\Omega_{k0}$ are very large
($\Omega_{k0}=0.02\pm 0.10$), so that plotting evolution diagrams
for statefinder parameters with confidence level errors might not
result in any useful information. Thus, we only discuss the best-fit
case, which is sufficient for our analysis. Using the best-fit
values, we plot in figure \ref{fig:obs} the statefinder diagrams
$r(s)$ and $r(q)$ for the spatially non-flat holographic dark energy
model. The coordinates of today in the holographic dark energy model
locate at: $(s_0, r_0)=(-0.102, 1.357)$ and $(q_0, r_0)=(-0.590,
1.357)$. For comparison, we also plot the statefinder diagrams of
the $\Lambda$CDM model in both spatially flat and non-flat cases in
this figure. The parameters for non-flat $\Lambda$CDM are also
$\Omega_{\rm m0}=0.29$ and $\Omega_{k0}=0.02$, and the parameter for
flat $\Lambda$CDM is $\Omega_{\rm m0}=0.27$. We see clearly that in
the $s-r$ plane the flat $\Lambda$CDM is shown as a fixed point $(0,
1)$ and the non-flat $\Lambda$CDM is exhibited as a vertical line
segment (very short in this plot) with present coordinate $(0,
1.02)$; in the $q-r$ plane the flat $\Lambda$CDM behaves as a
horizontal line segment with present coordinate $(-0.595, 1.00)$ and
the non-flat $\Lambda$CDM behaves as an arc with present coordinate
$(-0.585, 1.02)$. Note that the true values of $(s_0, r_0)$ of the
universe should be determined in a model-independent way, we can
only pin our hope on the future experiments to achieve this. We
strongly expect that the future high-precision experiments (e.g.
SNAP) may provide sufficiently large amount of precise data to
release the information of statefinders $\{H, q, r, s\}$ in a
model-independent manner so as to supply a way of discriminating
different cosmological models with or without dark energy.

\begin{figure}[htbp]
\begin{center}
\includegraphics[scale=1.0]{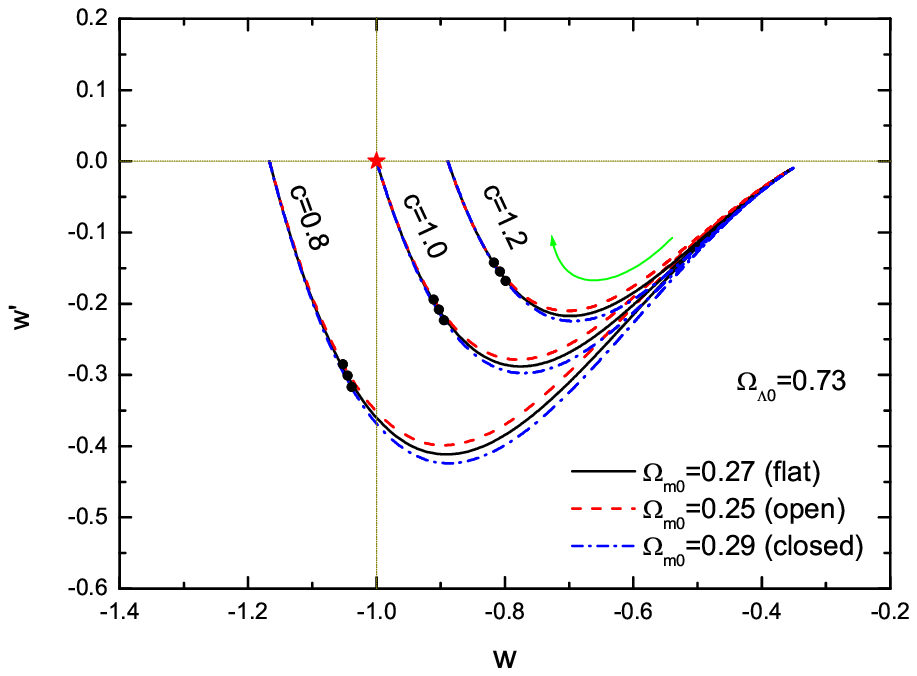}
\caption[]{\small The dynamical evolution behavior of the
holographic dark energy in the $w-w'$ plane. This diagram shows an
illustrative example in which we plot the evolution trajectories of
the holographic dark energy model in $w-w'$ plane by fixing
$\Omega_{\Lambda 0}=0.73$ and varying $\Omega_{\rm m0}$ as 0.27,
0.25 and 0.29 corresponding to the flat, open and closed universes,
respectively. Selected curves correspond to $c=1.2$, 1.0 and 0.8 for
including various representative cases. The arrow in the diagram
denotes the evolution direction. Dots denote the present values of
$(w, w')$. For comparison, the case for the cosmological constant
$\lambda$ is also marked in the $w-w'$ plane as a red star.}
\label{fig:wwp}
\end{center}
\end{figure}

\begin{figure}[htbp]
\centering $\begin{array}{c@{\hspace{0.2in}}c}
\multicolumn{1}{l}{\mbox{}} &
\multicolumn{1}{l}{\mbox{}} \\
\includegraphics[scale=0.8]{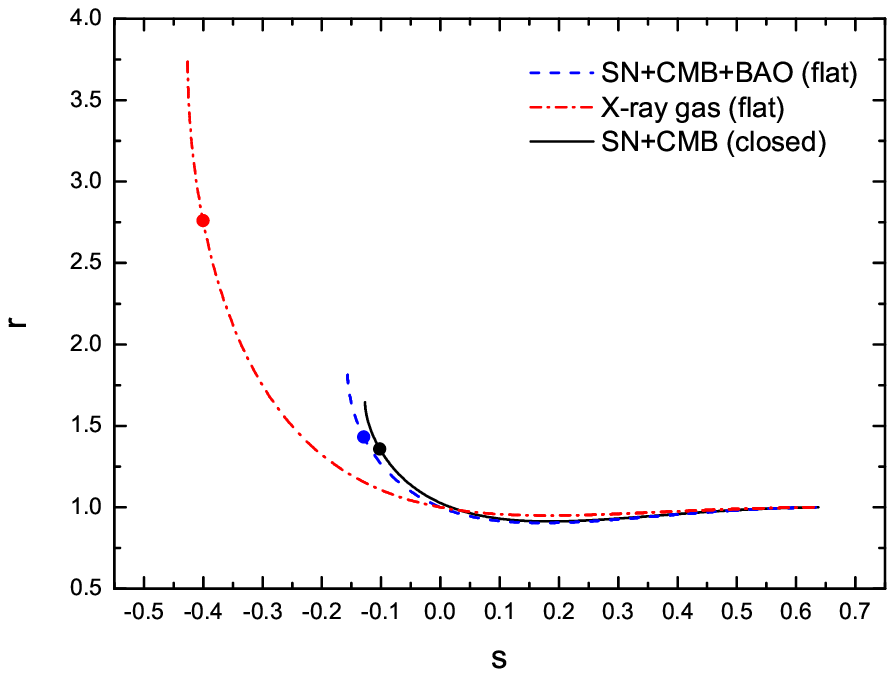} &\includegraphics[scale=0.8]{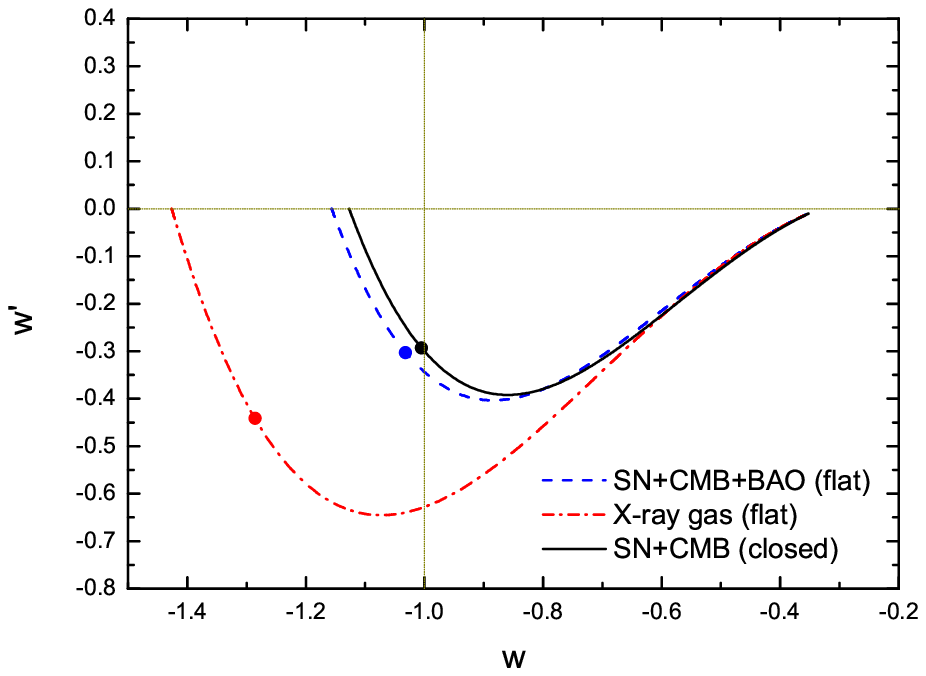} \\
\end{array}$
\caption[]{\small The $s-r$ diagram and the $w-w'$ diagram in three
cases of evolution of the holographic dark energy in flat and closed
universes corresponding to three best-fit results, SN+CMB+BAO and
X-ray gas in flat universe, and SN+CMB in non-flat universe. Dots
denote the present coordinates.} \label{fig:sfwwp}
\end{figure}

Also, it is of interest to discuss the dynamical property of the
holographic dark energy in the $w-w'$ phase plane. Recently,
Caldwell and Linder \cite{Caldwell:2005tm} proposed to explore the
evolving behavior of quintessence dark energy models and test the
limits of quintessence in the $w-w'$ plane, where $w'$ represents
the derivative of $w$ with respect to $\ln a$, and showed that the
area occupied by quintessence models in the phase plane can be
divided into thawing and freezing regions. Then, it became somewhat
popular for analyzing dark energy models in the $w-w'$ plane. The
method was used to analyze the dynamical property of other dark
energy models including more general quintessence models
\cite{Scherrer:2005je}, phantom models \cite{Chiba:2005tj} and
quintom models \cite{Guo:2006pc}, etc.. The $w-w'$ analysis
undoubtedly provides us with an alternative way of classifying dark
energy models using the quantities describing the property of dark
energy. But, it is obviously that the $(w, w')$ pair is related to
statefinder pair $(s, r)$ in a definite way, see Eqs. (\ref{rw}) and
(\ref{sw}). The merit of the statefinder diagnosis method roots that
the statefinder parameters are constructed from the scale factor $a$
and its derivatives, and they are expected to be extracted in a
model-independent way from observational data, although it seems
hard to achieve this at present. Now let us investigate the
holographic dark energy in the $w-w'$ phase plane. Figure
\ref{fig:wwp} shows an illustrative example in which we plot the
evolution trajectories of the holographic dark energy model in
$w-w'$ plane by fixing $\Omega_{\Lambda 0}=0.73$ and varying
$\Omega_{\rm m0}$ as 0.27, 0.25 and 0.29 corresponding to the flat,
open and closed universes, respectively. Selected curves correspond
to $c=1.2$, 1.0 and 0.8 for including various representative cases.
The arrow in the diagram denote the evolution direction. We see
clearly that the parameter $c$ plays a crucial role in the model:
$c\geqslant 1$ makes the holographic dark energy behave as
quintessence-type dark energy with $w\geqslant -1$, and $c<1$ makes
the holographic dark energy behave as quintom-type dark energy with
$w$ crossing $-1$ during the evolution history. As is shown in this
diagram that the value of $w$ decreases monotonically while the
value of $w'$ first decreases from zero to a minimum then increases
to zero again. We also notice that the effect of the spatial
curvature contribution can be identified explicitly in this diagram.
Furthermore, we compare predictions made by different fits of
observational data in the $s-r$ plane and the $w-w'$ plane. Figure
\ref{fig:sfwwp} shows the $s-r$ diagram and the $w-w'$ diagram in
three cases of evolution of the holographic dark energy in flat and
closed universes corresponding to three best-fit results, SN+CMB+BAO
and X-ray gas in flat universe, and SN+CMB in non-flat universe, as
mentioned above. So we see that the $w-w'$ dynamical diagnosis can
provide us with a useful complement to the statefinder $s-r$
geometrical diagnosis.


In summary, we have in this paper studied the holographic dark
energy model in a non-flat universe from the statefinder viewpoint.
Since the accelerated expansion of the universe was found by
astronomical observations, many cosmological models involving dark
energy component or modifying gravity have been proposed to
interpret this cosmic acceleration. This leads to a problem of how
to discriminate between these various contenders. The statefinder
diagnosis provides a useful tool to break the possible degeneracy of
different cosmological models by constructing the parameters $\{r,
s\}$ using the higher derivative of the scale factor. So the method
of plotting the evolution trajectories of dark energy models in the
statefinder plane can be used to as a diagnostic tool to
discriminate between different models. Furthermore, the values of
$\{r, s\}$ of today, if can be extracted from precise observational
data in a model-independent way, can be viewed as a discriminator in
testing the various cosmological models. On the other hand, though
we are lacking an underlying theory of the dark energy, this theory
is presumed to possess some features of a quantum gravity theory,
which can be explored speculatively by taking the holographic
principle of quantum gravity theory into account. So the holographic
dark energy model provides us with an attempt to explore the essence
of dark energy within a framework of fundamental theory. We perform
a statefinder diagnosis for the holographic dark energy model in a
non-flat universe in this paper. The statefinder diagrams show that
the contributions of the spatial curvature in the model can be
diagnosed out explicitly in this method. We hope that the future
high-precision observations such as the SNAP-type experiment may be
capable of providing large amount of accurate data to determine the
statefinder parameters precisely and consequently single out right
cosmological models.

\section*{Acknowledgements}

We would like to thank the referee for providing us with many
helpful suggestions. This work is supported in part by the Natural
Science Foundation of China.

\end{document}